\documentstyle[twocolumn,prl,aps]{revtex}
\begin{document}
\twocolumn[\hsize\textwidth\columnwidth\hsize\csname @twocolumnfalse\endcsname

\draft

\title{Landau Level Mixing and Levitation of Extended States
in Two Dimensions
}

\author{F. D. M. Haldane$^{1}$ and Kun Yang$^{1,2}$}
\address{$^{1}$Department of Physics, Princeton University,
Princeton, NJ 08544}
\address{$^{2}$Department of Electrical Engineering, Princeton University,
Princeton, NJ 08544}
\date{June 20, 1996}

\maketitle

\begin{abstract}
We study the effects of mixing of  different Landau levels on the 
energies of one-body
states, in the presence of a strong uniform magnetic field and a random
potential in two dimensions. 
We use a perturbative approach and develop a systematic
expansion in both the strength and smoothness of the random potential.  
We find the energies of the extended states
shift {\em upward}, and the amount of levitation 
is proportional to $(n+1/2)/B^3$ for strong magnetic field,
where $B$ is the magnetic field strength and $n$ is the Landau level index. 
\end{abstract}


]

\narrowtext
The behavior of extended electronic states of non-interacting
electrons in a uniform
magnetic field $B$  
with a random substrate
potential $V({\bf r})$, is of central importance to the understanding of the
integer quantum Hall effect\cite{review}.
In this Letter, we report a new and rather simple calculation that
exposes the microscopic origin of the so-called ``levitation''
of extended states\cite{lk,laughlin} in the large-$B$ limit, which has been the
subject of recent interest.

On the one hand, it is now widely accepted that, in the limit $B$ = 0,
there are no extended (delocalized) single-electron states at any finite
energy\cite{gangof4}, while in the strong field limit, there exist discrete 
energies near the center of each disorder-broadened Landau level,
at which states are extended.
An appealing (but heuristic) 
scenario, known as the ``levitation''
of extended states, has been proposed to explain how the interpolation
between these limiting behaviors might occur\cite{lk,laughlin}.
This holds that
one-electron states are localized at all energies except at a discrete
set $E_n^c (B)$ = $(n + \case{1}{2})\hbar \omega_c + \epsilon_n(B)$,
$n \ge 0$,
where $\omega_c$ = $|eB|/m$.   
The energies 
$\epsilon_n(B) \rightarrow \epsilon_c$, a constant\cite{note}, as
$|B| \rightarrow \infty $, and increase monotonically as $|B|$ decreases,
in such a way that $E^c_n(B) < E^c_{n+1}(B) < \infty $ for $|B| > 0 $, 
finally
diverging as $B \rightarrow 0 $. 

As the Fermi level is raised through $E^c_n(B)$, 
the $T=0$ Hall conductance jumps
from $ne^2/h$ to $(n+1)e^2/h$.  
As the Fermi energy approaches the critical energy 
from above or below, the
localization length of single-electron states at the Fermi level
diverges, and a zero-temperature quantum critical point is approached.
This has been widely studied in numerical 
simulations\cite{chalker,theory,huo,ld}.
Using the notion of a law
of corresponding states related by a Chern-Simons effective field theory,
this picture has been suitably reinterpreted and 
extended to interacting electrons and
the fractional quantum Hall effect\cite{klz}.   

While the levitation scenario is appealing, it has 
apparently not yet been derived from
microscopic considerations, and recently there has been
considerable interest in testing it experimentally and 
numerically\cite{a,glozman,krav,furneaux,shahbazyan,kag,liu,yang,sheng,gramada},
and in identifying its microscopic origin.  The  effect must be associated
with Landau-level mixing, which gives rise to an apparent paradox:
generically, mixing gives rise to a {\it level-repulsion} effect,
which would tend to {\it lower} rather than raise the energy levels.
(This is clear for the case $n = 0$, but is generally true, as the
level repulsion due to mixing with  higher Landau levels is always stronger
than that from lower ones.)

In this Letter, we resolve this
paradox, and provide what appears to be a rather simple derivation
of the initial appearance of levitation associated with Landau-level
mixing at large but finite fields, namely the $O(B^{-3})$ levitation
\begin{equation}
\epsilon_n(B) = \epsilon_c + (n+\case{1}{2})\hbar \omega_c
\left ( {(\ell/\xi)^2\over \omega_c \tau} \right )^2 + O(B^{-4}) ,
\label{result}
\end{equation}
where $\ell$ = $\sqrt{\hbar /|eB|}$ is the ``magnetic length'' (the
classical radius of the ground-state cyclotron orbit).
Here $\hbar/ \tau $ is the energy scale of Landau-level
broadening in the high-field limit (essentially the variance of
the fluctuations of $V({\bf r})$), and $\xi$ is a characteristic
length scale over which the potential varies by this amount.  This result
is derived in the limit $\ell / \xi  << 1$ and $\omega_c \tau >> 1$,
which is always achieved at sufficiently high magnetic fields
provided the potential is bounded, local, and smoothly-varying.

In the large-$B$ limit, when Landau quantization becomes exact,
the dynamics of cyclotron and ``guiding center" motion of electrons decouple, 
and the latter
can be treated semiclassically: the electrons
move adiabatically along equipotentials of the potential $V({\bf r})$,
with the local drift-velocity, ${\bf v}_d$ =
${\bf \hat z} \times \mbox{\boldmath$\nabla$} V ({\bf r})/eB $, 
where ${\bf \hat z}$ is the direction of the magnetic field
normal to the two-dimensional surface.  

If the topology of the region $V({\bf r}) < \epsilon $ is considered
as a function of $\epsilon$, Trugman\cite{trugman} noted that
there is a classical percolation transition
between a picture of disconnected ``lakes'' in a dry ``continent'' for
$\epsilon < \epsilon_c$, and a picture of a continuous ``ocean'' with
isolated dry ``islands'' for $\epsilon > \epsilon_c$.
Thus as the Fermi energy is raised, there is a transition in which
the regions where the Landau level is locally filled join up to
form a continuously-connected region, giving rise to a quantum Hall effect.

Corrections to this semiclassical behavior will
occur when the equipotential line on which a particle is moving
comes close to a saddle-point of $V({\bf r})$, and tunneling to a
nearby equipotential line at the same energy can occur\cite{fertig}. This
breakdown is believed to control the quantum critical behavior
when $\epsilon$ is close to $\epsilon_c$.
The Chalker-Coddington ``Network model''\cite{chalker} attempts to model this
by replacing the Hamiltonian by an effective model, representing
it as a network of saddle-points with energies close to $\epsilon_c$,
connected by essentially inert directional leads (the equipotential
lines).  A random scattering matrix describes the transition amplitudes
between the two incoming and two outgoing leads at each saddle-point.

The extended states are thus clearly identified with saddle-point energies.
In the limit of the strong-field limit of a finite system, a single
{\it critical saddle-point} will control the energy at which
transmission across the system first occurs.
This picture allows the identification
\begin{equation}
\epsilon_c = V({\bf r}_c),
\label{sadd}
\end{equation}
where ${\bf r}_c$ is the location of the {\it critical saddle-point}.
If the thermodynamic limit is first taken, one may anticipate that
the distribution of saddle-point energies is singular at $\epsilon_c$,
and there will be many critical saddle-points satisfying (\ref{sadd}).
In either case, the energy of delocalized states in
the strong-field limit must be associated
with saddle-point energies.

Our result follows from a systematic analysis of mixing between
Landau levels, computed perturbatively in the strong-field limit.
It is summarized as follows: the leading effect of mixing is accounted for
by working in the strong-field limit as before, but now
replacing the actual potential $V({\bf r})$ by a 
{\it locally-renormalized Landau-level-dependent effective potential}
\begin{equation}
V^{(n)}_{\rm eff}({\bf r}) =   V({\bf r}) + \sum_{m\ge 2} V_m^{(n)}({\bf r}),
\label{potl}
\end{equation}
where $V^{(n)}_m({\bf r})$ $\propto B^{-m}$ as $B \rightarrow \infty$.
The leading $O(B^{-2})$ correction is given by
\begin{equation}
V_2^{(n)}({\bf r})  = 
- {\ell^2 \over 2\hbar \omega_c } 
|\mbox{\boldmath$ \nabla $} V({\bf r} )|^2 \le 0,
\end{equation}
which is independent of the Landau level index, and negative.
This is the generically-dominant  ``level-repulsion'' term, 
which indeed causes a downward shift of typical energy levels.
It is proportional to the square of the local electric field strength,
and is the only correction in the trivially-solvable case where
the substrate potential $V({\bf r})$ is that of a uniform electric field.

The crucial observation is that $\epsilon_n$ is given by
$V^{(n)}_{\rm eff} ({\bf r}_c)$, where ${\bf r}_c$ is the location
of the {\it critical saddle-point}. However, 
$\mbox{\boldmath$\nabla$}V({\bf r}_c)$ = 0:
this implies that the dominant correction $V_2^{(n)}({\bf r})$ 
{\em does not} affect the extended state energies.

The next order correction is
\begin{equation}
V_3^{(n)}({\bf r})  =
 \case{3}{8} (n +\case{1}{2})\left ( {\ell^4 \over
\hbar \omega_c } \right ) u({\bf r}),
\end{equation}
where
\begin{eqnarray}
u({\bf r}) &=&
 (\nabla^2 V({\bf r}))^2 - \det_{ij} | \nabla_i\nabla_j V({\bf r}) |, 
\nonumber
\\
&\equiv&
(\nabla_x^2 V- \nabla_y^2V)^2 + (2\nabla_x\nabla_yV)^2 \ge 0.
\label{end}
\end{eqnarray}
At a saddle-point
$u({\bf r}_c) > 0$, as the determinant of second derivatives
is negative.
(Note that
$u({\bf r})$ 
only vanishes if the matrix of second derivatives
is rotationally-invariant, which is not true at a saddle-point).
Thus, in contrast to a generic point where the corrections to the
effective potential due to Landau-level mixing are negative, the
leading correction at saddle-points, which control the energies
of extended states, is {\em positive}, giving rise to the levitation effect.
The result (\ref{result}) follows from an estimate of
$u({\bf r}_c)$ as being of order
$(\hbar/  \tau \xi^2 )^2$. 
Our results are schematically illustrated in Fig. (\ref{fig1}).

We now sketch  the technical derivation of
(\ref{potl})-(\ref{end}).
We write the substrate potential $V({\bf r})$ in terms
of its Fourier components $\tilde V ({\bf q})$
\begin{equation}
V({\bf r}) = {1 \over A} \sum_{\bf q} \tilde V({\bf q}) 
e^{i{\bf q}\cdot {\bf r}}
\end{equation}
where for convenience we have imposed
(quasi-)periodic boundary conditions on an area A that contains an
integral number of magnetic flux quanta.
We now write 
\begin{equation}
e^{i{\bf q}\cdot {\bf r}}
=
e^{i{\bf q}\cdot {\bf R}} U({\bf q}),\quad
U({\bf q}) = e^{i{\bf q} \cdot ({\bf r} - {\bf R})},
\end{equation}
where ${\bf R}$ is  the ``guiding center'' of the cyclotron orbit\cite{duncan},
which obeys the algebra\cite{gj,duncan}
\begin{equation}
e^{i {\bf q} \cdot {\bf R}}e^{ i {\bf q}' \cdot  {\bf R}}
= \exp (\case{1}{2}i ({\bf q} \times {\bf q}')\ell^2 )
e^{i ({\bf q}+{\bf q}') \cdot {\bf R}}.
\label{alg}
\end{equation}
(Here ${\bf q} \times {\bf q}'$ $\equiv$ $q_xq'_y - q_yq'_x$ .)
The unitary operator $U({\bf q})$ acts entirely on the
cyclotron orbit (Landau level) variables, and commutes with the guiding center.
In the strong-field limit, the potential term projected into the Landau
level $n$ becomes
\begin{equation}
{1\over A} \sum_{\bf q} \tilde V ({\bf q}) e^{i {\bf q}\cdot {\bf R} }
 U({\bf q})_{nn} ,
\label{ll}
\end{equation}
where $U ({\bf q})_{nn'}$ $\equiv$
$\langle n | U({\bf q}) |n' \rangle $ ($n$ and $n'$ are Landau-level indices):
for $n \ge n'$, $U_{nn'}({\bf q})$ is given by
\begin{equation}
\left ( { (q_x + i q_y)\ell \over \surd 2}\right )^{n-n'}
L^{n-n'}_{n'}(\case{1}{2}q^2\ell^2)
\exp (- \case{1}{4} q^2\ell^2), 
\label{lag}
\end{equation}
where $L^m_n(x)$ is a Laguerre polynomial.

The problem in the high-field limit is to diagonalize the projected
potential (\ref{ll}), in the subspace of a given Landau level.
When the field strength is strong but finite, 
states in different Landau
levels are still {\em well separated} in energy. Nevertheless,
electrons in a given Landau level may be
scattered into other Landau levels by the random potential, and will
eventually come back due to energy conservation.
The effect of such (virtual) processes is to
{\em renormalize} the {\em effective} potential seen by the
electrons in this Landau level [see Fig. (\ref{fig2})],
which we  calculate below.

The trick we
will use to characterize the renormalization is to develop a
perturbative expansion 
in $V/\hbar \omega_c$, and rewrite the effective Hamiltonian in the
form (\ref{ll}), but with a renormalized 
$\tilde V^{(n)}_{\rm eff}({\bf q})$, which can then  be expanded in powers
of $\ell$ as well as in $1/\hbar \omega_c$, to give
a true $1/B$ expansion. 
We then carry out the Fourier transform to find the renormalized
$V_{\rm eff}^{(n)}({\bf r})$ that this corresponds to.

Using standard perturbative renormalization formalism\cite{anderson},
we find the leading $O(V^2/\hbar \omega_c)$ term in the effective Hamiltonian
is
\begin{equation}
{1\over A^2} \sum_{{\bf q}{\bf q}'} {\tilde V ({\bf q})
\tilde V({\bf q}') \over \hbar \omega_c }
e^{i {\bf q}\cdot {\bf R} }e^{i {\bf q}'\cdot {\bf R} }
{\sum_{n'}}' { U_{nn'}({\bf q}) U_{n'n} ({\bf q'}) \over (n-n')}.
\end{equation}
The primed sum means that the singular term
$n' = n$ is excluded.
We must now express this term in the form (\ref{ll}),
using the contraction (\ref{alg}).   The general 
$O(V^m/(\hbar \omega_c)^{m-1})$ contribution
to $V_{\rm eff}^{(n)}({\bf r})$ may be written (for $m > 1$) 
in the form
\begin{equation}
{\hbar \omega_c \over A^m}\sum_{{\bf q}_1\ldots {\bf q}_m} 
\left ( \prod_{i=1}^m 
{ \tilde V({\bf q}_i) e^{i{\bf q}_i \cdot {\bf r}} \over \hbar \omega_c }
\right ) 
 f^{(n)}_m({\bf q}_1,\ldots , {\bf q}_m) ,
\end{equation}
where $f_m^{(n)}({\bf q}_1,\ldots,{\bf q}_m)$ is a {\it symmetric} 
and {\it analytic}
function of the $\{ {\bf q}_i\ell\}$
(it is derived from the $U_{nn'}({\bf q})$, which are analytic).  
It is also {\it rotationally-invariant},
and must {\it vanish as any of the ${\bf q}_i \rightarrow 0$}, as addition
of a
spatially-constant term (a ${\bf q} = 0$ Fourier component) 
to the
potential cannot affect the non-linear terms in 
$V^{(n)}_{\rm eff} ({\bf r}$).
The term $f^{(n)}_2 ({\bf q}_1,{\bf q}_2)$ is the symmetric part of
\begin{equation}
{e^{i {\bf q}_1\times {\bf q}_2 \ell^2/2} \over 
U_{nn}({\bf q}_1 + {\bf q}_2)}
{\sum_{n'}}' {U_{nn'}({\bf q}_1)U_{n'n}({\bf q}_2) \over 
(n-n')}.
\label{order2}
\end{equation}
It is straightforward to expand $f_2^{(n)}({\bf q}_1,{\bf q}_2)$ in powers
of $\ell$, using (\ref{lag}); 
we find that, up to terms of order $\ell^4$, it is given by
\begin{equation}
\case{1}{2}({\bf q}_1\cdot {\bf q}_2)\ell^2
+ \case{3}{8} (n+ \case{1}{2})
\left (  ( {\bf q}_1\cdot {\bf q}_2)^2 - ({\bf q}_1 \times {\bf q}_2)^2 \right )
\ell^4.
\end{equation}
This corresponds to a gradient expansion of the effective potential
in real space, and
gives the leading terms of $O(B^{-2})$ and $O(B^{-3})$ in 
(\ref{potl}). 

We find that the leading term in the gradient expansion
of the term of order $O(V^3/(\hbar\omega_c)^2)$
is of order $\ell^4$ (this  in fact follows directly from
the general properties of $f^{(n)}_3({\bf q}_1,{\bf q}_2,{\bf q}_3)$
mentioned above).   This means that its leading contribution
to the effective potential is $O(B^{-4})$,
and it does not contribute to the leading
terms. Higher-order terms in $V/\hbar \omega_c$ vanish even faster at large
$B$.

In the following we discuss the experimental implications of our results,
and the relation between these results and existing work on this subject.
The results above have interesting implications for attempts
to experimentally detect levitation based on the relative motion of
the extended state energy and the mean energy  of the broadened Landau
level (as defined using the density of states)
at large $B$. 

Our result shows that the leading effect
in this limit is an $O(B^{-2})$ {\it downwards} motion of the
mean energy of the Landau level, while the extended state is
{\it static}
to this order, and only levitates to $O(B^{-3})$. In this limit
at least, experimental evidence\cite{glozman,krav,furneaux}
that the extended state rises relative
to the mean energy of the Landau level would be  demonstrating
not levitation of extended states,
but the lowering of localized state energies due to level-repulsion between
Landau levels. We also note that evidence of levitation of extended states
has been found in previous numerical work, in both the continuum
system\cite{a,ld}, and the tight binding model\cite{yang}, although
there is controversy in the latter case\cite{liu,sheng}.

Recently Shabazyan and Raikh\cite{shahbazyan}
(see also Ref.\onlinecite{kag}) used an extension of the
network model\cite{chalker}
to simulate the continuum system in the presence of a smooth random
potential. They considered the effects
of strongly-localized orbitals of different
Landau levels with energies close to the saddle-point energies of a particular
Landau level of interest,
and find that resonant tunneling into such orbitals results 
{\em on average} in a reduction
of the transmission rate through the saddle-points, 
implying an upward shift of the energy of extended states,
which does not depend on the Landau level index $n$. We note
that in order for this effect to be important, there must be significant
overlap in the density of states (DOS) of different Landau levels; while it
is clear from our results that levitation occurs even if there is no
overlap in the DOS of different Landau levels (which is the case when $B$ is
large).

Later, Gramada and Raikh\cite{gramada} studied the effects of a short-range
impurity potential on the transmission rate through a nearby saddle-point,
and again find a reduction of the transmission rate {\em on average}.
They estimate the upward shift of the extended state energy due to this
effect to be of order $B^{-4}$ for large $B$.
We believe the $O(1/B^3)$ levitation we identify here is
the dominant one, at large $B$.

There are recent observations\cite{krav,shahar} of apparently-direct
transitions from quantum Hall states with large $\nu$ to insulating states
 at very
{\em weak} magnetic field, which appears to be inconsistent with the 
conventional one-electron  extended-state-levitation
picture and the global phase 
diagram\cite{klz}. We note at very weak magnetic field electron-electron
interactions may become important and the one-body picture 
may not be sufficient.  However, a quantitative validation of
the levitation scenario 
for  non-interacting electrons
in the $B \rightarrow 0 $ limit clearly urgently needs to be attempted.

To summarize: 
we have used a perturbative approach to study the effects of
mixing between Landau levels 
in a two-dimensional non-interacting electron system,
due to a random substrate potential. 
In high magnetic fields,
we find that although 
most of the states in the Landau level with index $n$
are pushed to lower energy 
by such mixing,
the energy of extended states shifts upward, 
and 
the amount of this shift is 
proportional to $(n+1/2)/B^3$.

We thank R. N. Bhatt, H. A. Fertig, S. M. Girvin, D. Shahar and S. L. Sondhi
for helpful discussions. This work was supported by NSF grant DMR-9400362.

\begin{figure}
\caption{Density of states and energy of extended states
in a given Landau level
before (dashed lines) and after (solid lines) Landau level 
mixing is taken into account.  
}
\label{fig1}
\end{figure}

\begin{figure}
\caption{Schematic perturbative expansion of the effective potential
seen by electrons in the $n$th Landau level.
}
\label{fig2}
\end{figure}
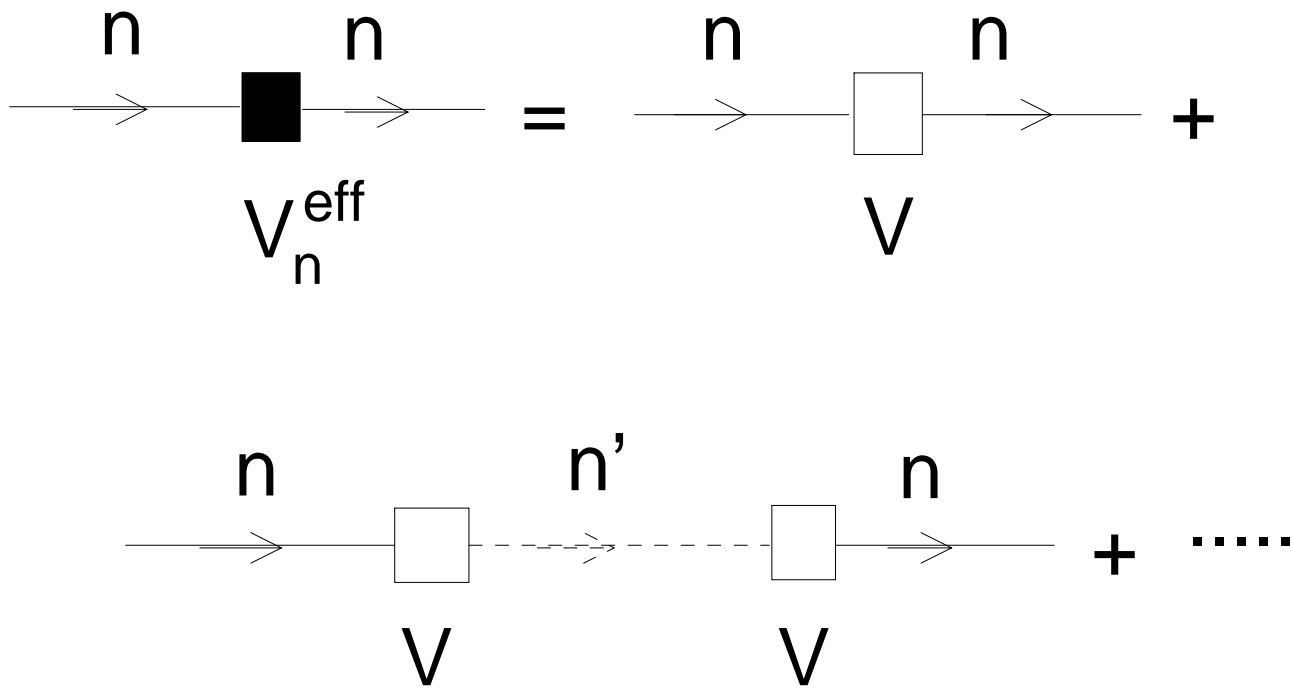
\end{document}